# Valley-dependent topological edge states in plasma photonic crystals

Jianfei LI[1], Chen ZHOU[1], Jingfeng YAO[1,2,*], Chengxun YUAN[1,2,*], Ying WANG[1,2], Zhongxiang ZHOU[1,2], Jingwen ZHANG[1]  and Anatoly A. KUDRYAVTSEV[3]

[1] School of Physics, Harbin Institute of Technology, Harbin 150000, People's Republic of China
[2] Heilongjiang Provincial Key Laboratory of Plasma Physics and Application Technology, Harbin 150000, People's Republic of China
[3] Physics Department, St. Petersburg State University, St. Petersburg 198504, Russia
**Email**: yaojf@hit.edu.cn and yuancx@hit.edu.cn

**Abstract**

Plasma photonic crystals designed in this paper are composed of gas discharge tubes to control the flow of electromagnetic waves. The band structures calculated by the finite element method are consistent with the experimental results in that two distinct attenuation peaks appear in the ranges of 1 GHz to 2.5 GHz and 5 GHz to 6 GHz. Electromagnetic parameters of the plasma are extracted by the Nicolson-Ross-Weir method and effective medium theory. The measured electron density is between $1\times10^{11}\,\text{cm}^{-3}$ and $1\times10^{12}\,\text{cm}^{-3}$, which verifies the correctness of the parameter used in the simulation, and the collision frequency is near $1.5\times10^{10}\,\text{Hz}$. As the band structures are corroborated by the measured scattering parameters, we introduce the concept of photonic topological insulator based on the quantum Valley Hall effect into the plasma photonic crystal. A valley-dependent plasma photonic crystal with hexagonal lattice is constructed, and the phase transition of the valley $K$ ($K'$) occurs by breaking the spatial inversion symmetry. Valley-spin locked topological edge states are generated and excited by chiral sources. The frequency of the non-bulk state can be dynamically regulated by the electron density. This concept paves the way for novel, tunable topological edge states. More interestingly, the Dirac cone is broken when the electron density increases to $3.1\times10^{12}\,\text{cm}^{-3}$, which distinguishes from the methods of applying a magnetic field and changing the symmetry of the point group.

Keywords: plasma photonic crystal, valley-dependent, topological state, electron density

## 1. Introduction

Plasma is widely used in stealth applications, broadband absorber, and plasma antenna fields by virtue of its extraordinary physical properties such as complex permittivity, electrical conductivity, and reconfigurability[1–3]. The electron density and collision frequency substantially affect the reflection, absorption, and refraction of electromagnetic waves when electromagnetic waves propagate in plasma. The combination of plasma and metamaterials shows promising applications in dynamically controlling the flow of electromagnetic waves. H. Hojo first proposed the concept of plasma photonic crystals, arranging the plasma and quartz periodically in one dimension. The frequency gap and cut-off band were calculated by solving the Maxwell equations, and the transmittance profile could be modulated by the electron density[4,5]. This successfully exploited the merits of plasma and dielectric photonic crystals. Because of the tunable and reconfigurable properties of plasma, the study of plasma photonic crystals has been extended from microwave to terahertz bands, and the research dimension has also been extended from one-dimension to three-dimension theoretically[6,7]. In the experiment, O. Sakai constructed two-dimensional plasma photonic crystals in experiments that included plasma columns generated by the dielectric barrier discharge and the background air. The transmittance had a clear attenuation peak at 73.83 GHz when the transverse electric (in-plane electric field) mode was excited, which was consistent with the results of theoretical calculations using the plane-wave expansion method[8,9]. In addition, they designed plasma metamaterials where double metallic helices were inserted in the waveguide periodically to produce negative permeability. At the same time, the metal helices were applied with low-frequency high voltage to produce microplasma (i.e., negative permittivity at specific frequencies). The combination of the two designs produced a double negative refraction property[10]. A tunable negative-refractive-index device was designed by M. A. Cappelli, which consisted of plasma discharge tubes and double splitring resonators. Negative permittivity and permeability were achieved in the range of 2 GHz to 2.47 GHz, and the transmittance increased with increasing current over a specific range[11]. F. Liu et al. designed a plasma photonic crystal with annular lattices, a self-organized form generated by dielectric barrier discharge. The transition from annular lattice to core-annular and concentric-annular lattice was regulated by the applied voltage, and the forbidden band between 28 GHz to 30 GHz was verified experimentally[12]. W. Fan et al. used dielectric barrier discharge to form honeycomb plasma photonic crystal. The lattice structure can be regulated by the applied voltage, which is a form of self-organization[13]. In the previous study of our group, we used dielectric columns and a plasma background to form a plasma photonic crystal with a square lattice. By dynamically modulating the electron density of the plasma, a triple degenerated Dirac cone was obtained at the centre of the Brillouin zone. When electromagnetic waves are incident at the frequency of the Dirac point, the plasma photonic crystal has zero permittivity and permeability[14].

With the successful development of photonic crystals, the photonic topological insulator analogous to electronic systems has attracted extensive research. Topologically-protected edge modes based on the quantum Hall effect and the quantum-spin Hall effect have promising applications in the field of waveguides, coupled resonators, chip development, and biosensors[15–17]. Besides, the topological effect can also be achieved by adjusting the valley degrees of freedom of the valley photonic crystal, which has been proposed in recent years. By breaking the spatial inversion symmetry, topological phase transitions arise at the $K$ ($K'$) valley in the Brillouin zone, and valley-spin locked topological edge states were detected theoretically and experimentally[18–20]. However, topological edge states in all-dielectric photonic crystals can only be observed in a specific frequency range with fixed lattice structures and electromagnetic parameters, and the electromagnetic modes are not switchable. Several materials such as liquid crystals and barium titanate have been used to explore the reconfigurable topological states[21,22]. P. Qiu et al. used plasmonic to form honeycomb lattices. Topological angular states were verified by changing the radius of the column or the position of the plasmonic[23]. Y. Zhao et al. filled nematic liquid crystals into dielectric columns to construct valley photonic crystals while topological phase transitions were produced by changing the position of the liquid crystal columns. In theory, the anisotropic permittivity of liquid crystals can be used to modulate and encode topological states [24]. Meanwhile, silicon columns and liquid crystal backgrounds were combined into valley photonic crystals designed by M. I. Shalaev, which realized the modulation of topological states by changing the anisotropic permittivity of liquid crystals[25].

In this paper, we investigate valley-dependent topological edge states in plasma photonic crystals composed of gaseous plasma, dielectric layers, and air background. Firstly, a two-dimensional plasma photonic crystal is constructed with discharge tubes, and the transmittance of electromagnetic waves is measured to verify the simulated band structures, which confirms the feasibility of constructing photonic crystals using gaseous plasma. Next, we test the scattering parameters (S-parameters) of the discharge plasma within the Transverse Electro-Magnetic (TEM) cell and extract the electromagnetic parameters via the NRW method and effective medium theory. Finally, the tunable topological edge states are realized theoretically by constructing plasma valley photonic crystals.

**2. Plasma photonic crystal with square lattice**

Plasma photonic crystals enable the modulation of the electromagnetic wave since the electromagnetic parameters can be dynamically controlled by external conditions such as applied voltages, gas pressure and magnetic fields. At the same time, the band structures of plasma photonic crystals can be calculated precisely as in the case of all-dielectric photonic crystals. In particular, it is necessary to accurately acquire the electromagnetic parameters of plasma. In plasma physics, a common way to generate plasma is capacitively coupled discharge. The fluorescent lamp provides beneficial conditions for discharge plasma as it is filled with an Ar-Hg mixture, and the pressure is 3 torr. A two-dimensional plasma photonic crystal with square lattice is composed of 28 lamps. The experimental setup is shown in figure 1(a), and the primitive cell is a square lattice with a lattice constant of 30 mm as shown in the inset. Two horn antennas with 1 to 14 GHz placed on both sides of the plasma photonic crystal to measure the S-parameters, and the measurement system is surrounded by wave-absorbing sponges.

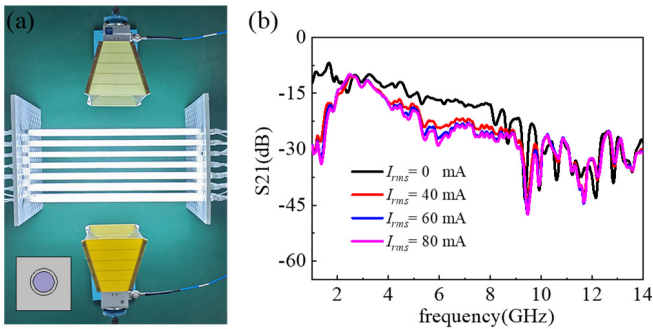

Figure 1. (a) The experimental setup of a plasma photonic crystal with square lattices; (b) The measured S21 parameters at different discharge currents.

All the lamps are connected in series to ensure that each lamp has the same discharge state and produces the same electron density. It is known that the lamp has a low maintained voltage but a high excited voltage. The power source of CTP-2000K is used to support the discharge of the lamps. The pulsed high frequency oscillator is adopted to generate high voltages of 15 kV and very low currents. Then all the lamps can be ignited by the pulsed high frequency oscillator, and the discharge current is regulated by the power source. It must be noted that the power source line needs a diode in series to isolate the high voltage generated by the pulsed high frequency oscillator. The measured results at different discharge currents are shown in figure 1(b), two distinct attenuation peaks appear in the range of 1 GHz to 2.5 GHz and 5 GHz to 6 GHz, and the magnitude of attenuation increases with increasing discharge current. However, in the high-frequency region, the attenuation peak shifts rather than changes in magnitude at different currents, which verifies the existence of two forbidden bands in the plasma photonic crystal.

Next, the experiments to precisely measure the plasma electron density and collision frequency are performed in the TEM cell. The Nicolson-Ross-Weir (NRW) method is used to extract the electron density and collision frequency from the measured scattering parameters[26,27]. This approach of electromagnetic wave diagnosis does not require the electron kinetic temperature, the electron energy distribution function, or the collision cross section[28]. figure 2(a) shows the top view of the measuring apparatus, which consists of a TEM cell and eight mercury-based fluorescent lamps. The length of the TEM cell is 430 mm, and the height is 60 mm. At the centre of the TEM cell, the transverse electromagnetic wave is formed with a length of 152 mm[29,30]. Eight fluorescent lamps with an outer radius of 6.33 mm and an inner radius of 5.6 mm are neatly arranged in the central area of the TEM cell, and the energy flow of electromagnetic waves is perpendicular to the sample as shown in figure 2(b). The red line represents the electric field and the blue line represents the magnetic field. So, the effective medium theory base on the Garnett rule can suitably extract the complex permittivity and permeability of plasma[31,32]. The effective sample is shown as the black dashed line in the figure, which consists of plasma, quartz tube, and air. The volume ratio of the plasma to the effective sample is $f = 0.5188$. NRW method is commonly used in testing electromagnetic parameters of the sample. In this paper, the TEM cell can be viewed as a coaxial transmission line which accurately measure scattering parameters in a closed field compared to free space. According to Nicolson's algorithm, the permittivity and permeability of the sample can be derived from the measured S-parameters[26,27]:

$$\mu_r(\omega) = \frac{2\pi\omega}{\lambda_l c} \cdot \frac{1+\Gamma}{1-\Gamma} \quad (1)$$

$$\varepsilon_r(\omega) = \frac{(2\pi c)^2}{\omega^2 \lambda_l \mu_l} \quad (2)$$

where reflection coefficient is $\Gamma = X \pm \sqrt{X^2-1}\,(|\Gamma| \leq 1)$, and the relations between reflection coefficient and scattering parameters is $X = \left[1-\left(S_{21}^2 + S_{11}^2\right)\right]/2S_{11}$. $\mu_r$, $\varepsilon_r$, $l$, $\omega$, $\lambda_l$ are permeability, permittivity, length of the sample, radian frequency of the incident wave, and the wavelength in the sample, respectively. In the experiment, eight lamps are connected in series to ensure the same discharge state of each fluorescent lamp and are driven by the CTP-2000K.

The measured S21 is shown in figure 2(c). When the root mean square of the current reaches 38 mA, all lamps are discharged. However, S21 is attenuated by 20 dB on average as shown in the green line, which is not only the effect of the plasma but more importantly the mercury vapor. The extracted electron density and collision frequency from this parameter do not truly reflect the properties of the plasma. This leads the collision frequency calculations that are too large, such as 29 GHz to 74 GHz, as seen in the literature[33]. After that, the attenuation of S21 gradually increases with the increasing current. It is worth noting that the discharge current increases to 128 mA and a distinct cut-off region appears at less than 5 GHz, in which the permittivity of plasma is negative, as shown in the left part of the red line. As seen in figure 2(d), S11 is almost the same when the $I_{rms} = 64$ mA and $I_{rms} = 128$ mA, which means that the collisional absorption of electromagnetic waves by the plasma is weakened. Then, the effective medium theory is combined with the NRW algorithm to accurately diagnose the plasma parameters. To exclude the effect of mercury vapor, the complex permittivity of the effective sample is used as the background medium when the current is 38 mA. The classical Maxwell Garnett approximation for extracting the complex permittivity of plasma is adopted as follows[32],

$$\varepsilon_{eff} = \varepsilon_2 \left[1 + 3f\frac{\varepsilon_1-\varepsilon_2}{\varepsilon_1+2\varepsilon_2} \bigg/ \left(1-f\frac{\varepsilon_1-\varepsilon_2}{\varepsilon_1+2\varepsilon_2}\right)\right] \quad (3)$$

where $\varepsilon_{eff}$ is the total effective permittivity of the sample extracted from the S-parameter using the NRW method, $\varepsilon_1$ is the permittivity of plasma, $\varepsilon_2$ is the permittivity of background medium, and $f$ is the volume ratio of the plasma to the background medium (in this case $f = 0.5188$). The measured S-parameters are used as a background medium to eliminate the effect of mercury vapor when $I_{rms} = 38$ mA, and the complex permittivity of the plasma is obtained.

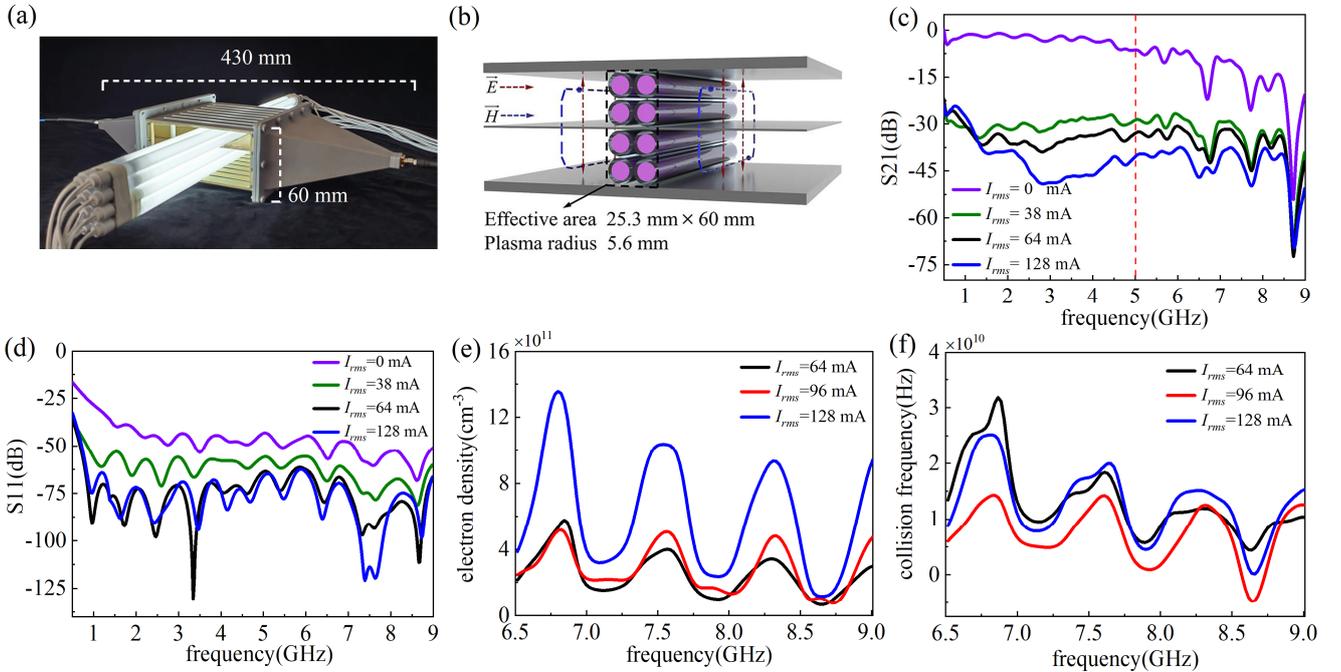

Figure 2. (a) Experimental diagram of the S-parameter measurement in the TEM cell; (b) A schematic of the effective sample in the TEM cell; (c) The measured S21 over a wide frequency ranges; (d) The measured S11 over a wide frequency range; (e) The electron density

calculated by the NRW method and effective medium theory; (f) The collisional frequency calculated by the NRW method and effective medium theory.

Then, the electron density and collision frequency of plasma can be derived from Drude equation:

$$\varepsilon(\omega) = 1 - \frac{\omega_{pe}^2}{\omega(\omega - i\nu_c)} \quad (4)$$

where $\omega_{pe}$ is the Langmuir frequency and $\nu_c$ is the electron-neutral collision frequency. The Langmuir frequency can be denoted as $\omega_{pe} = \sqrt{(ne^2/\varepsilon_0 m_e)}$, where $n$ is the electron density, $e$ is the unit charge, and $m_e$ is the mass of the electron.

In the previous literatures, the analysis of the collision frequency is ambiguous when electromagnetic waves propagate in the plasma generated by fluorescent lamps. Here, we can precisely extract the plasma parameters using effective medium theory. The plasma cut-off frequency is around 5 GHz due to the low transmittance in the region less than that frequency. From figure 2(e), it can be seen that the electron density increases with current. When the current reaches 128 mA, the electron density is around $8 \times 10^{11}$ cm$^{-3}$. The collision frequency is around 15 GHz as shown in figure 2 (f). It is found that the electron density and collision frequency of the plasma fluctuate with the radian frequency of the incident wave. Because the measured S-parameter consists of a real and an imaginary part. For example, the real part of S21 is $E_2/E_1 \cos(\phi_2 - \phi_1)$ and the imaginary part of S21 is $E_2/E_1 \sin(\phi_2 - \phi_1)$ where $\phi_n$ is the phase that contains the radian frequency. The obtained permittivity and permeability of plasma must fluctuate with frequency. Therefore, the extracted electron density and collision frequency fluctuate in a small range with the frequency of incident wave.

The calculation of the dispersion relationship is extremely valuable and can accurately predict the transport properties of electromagnetic waves. The critical issue is that experimental results are necessary to validate the simulation. It is reasonable to ignore the collision frequency in substance because the electromagnetic energy will be concentrated at the high permittivity region when the electromagnetic wave propagates in photonic crystals[34]. The band structure of the plasma photonic crystal is calculated using the finite element method (FEM) by COMSOL software. Before discharge, the band structure of the photonic crystal composed of fluorescent lamps is shown in the left part of figure 3. There is no forbidden band in the Γ-X direction which is the direction of electromagnetic wave propagation in the experiment. When the plasma is added to the lamps, the calculation of the band structure becomes complicated due to the dispersive properties of plasma. The eigenfrequency interface and global ODE and DAE interfaces must be combined to find the eigenvalues in the primitive cell.

Here we use the collisionless cold plasma approximation whose permittivity is expressed as $\varepsilon_r(\omega) = 1 - \omega_{pe}^2/\omega^2$, and the plasma electron density of $5 \times 10^{11}$ cm$^{-3}$ is adopted. The calculation results are shown in the right part of figure 3. It is obvious to see two forbidden bands in the range of 0 GHz to 1.9 GHz and 4.9 GHz to 5.5 GHz, which is highly consistent with the experimental results. An omnidirectional forbidden band appears in the low-frequency band, which is a typical feature of plasma photonic crystals and has been extensively studied in the literature[9,35–37]. The collisionless cold plasma approximation is reasonable through theoretical and experimental studies of plasma photonic crystals with a square lattice. These results show that the plasma photonic crystal composed of gaseous plasma has a distinct band gap characteristic, which will provide the basis for tunable devices based on plasma. Meanwhile, the novel effects in optics such as topology, pseudo-diffusive transport, and zero refraction have yet to be investigated in plasma photonic crystals. Next, we will focus on topological edge states based on the valley Hall effect.

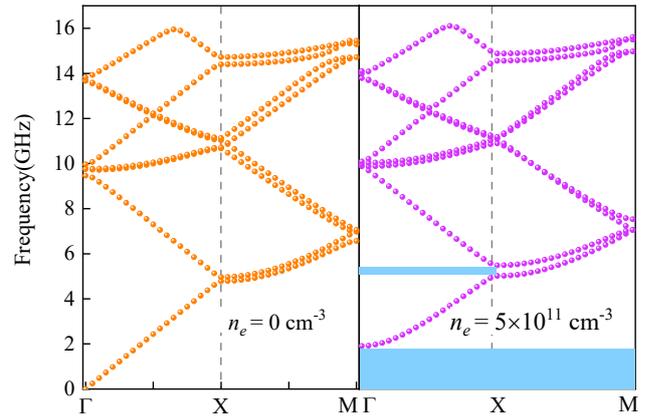

Figure 3. Band structures of plasma photonic crystal calculated by the finite element method at different electron densities.

## 3. Valley photonic crystal based on gaseous plasma

In the previous section, we obtained the electron density in the fluorescent lamps by the NRW method and demonstrated that the collisionless cold plasma approximation is reasonable through S21 in plasma photonic crystal. Then, the valley plasma photonic crystal is designed to realize the topological edge states, which possess dynamically tunable properties. We design the plasma photonic crystal with the honeycomb lattice as shown in figure 4(a). The grey area is the primitive cell with a lattice constant of $a = 30$ mm . Each column consists of a ring-shaped solid dielectric material with a permittivity of 11.7, whose inner radius is $0.154a$ and outer radius of $0.2a$ , and gaseous plasma.

They are distributed periodically in the x and y directions, and the length is infinite in the z direction. The electron density used in the simulation is $8\times10^{11}$ cm$^{-3}$ . Columns of types 1, 2, and 3 are connected in series, and columns of types 4, 5, and 6 are connected in series, which ensures that they can be controlled separately. In this work, the TM mode of electromagnetic wave with out-of-plane electric fields is considered, and the band structure of plasma photonic crystal is calculated as shown in figure 4(b).

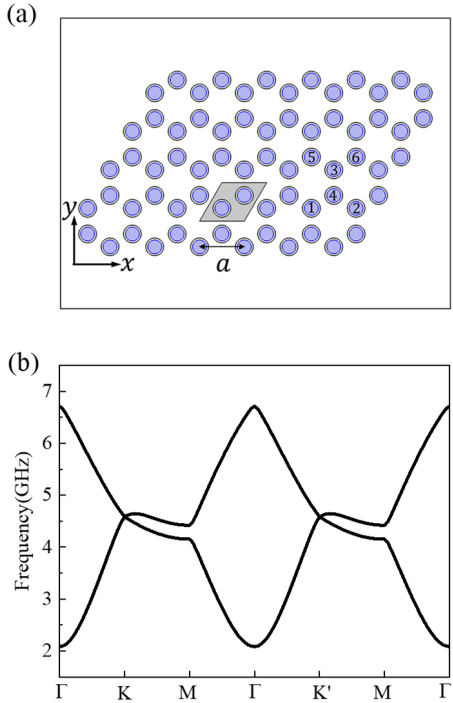

Figure 4. (a) The plasma photonic crystal with a honeycomb lattice; (b) The calculated band structure at $8\times10^{11}$ cm$^{-3}$ .

The $C_{3v}$ symmetry and time-reversal symmetry are preserved when all tubes are filled with gaseous plasma, and the Dirac cone is formed at the $K$ valley and $K'$ valley in the Reduced Brillouin zone of plasma photonic crystal. The horizontal axis represents the wave vector in the reciprocal lattice space. For the calculation, the wave vector $k$ is scanned at the high symmetry points of the reduced Brillouin zone to obtain the eigenfrequencies. For more intuitive results, the vertical axis uses frequency units rather than the normalized frequency $\omega a/2\pi c$ where $\omega$ is the radian frequency and $c$ is the light speed. When the dispersive material is introduced into the calculation of the band structures, an omnidirectional band gap appears in the entire Brillouin zone below 2.08 GHz.

Valley cleavage occurs when the spatial inversion symmetry is broken, whose lower left column or upper right in the primitive cell are filled with gaseous plasma, as shown in figure 5(a) and (c). In this case, an omnidirectional bandgap (4GHz to 4.3GHz) arises here because the symmetry of the honeycomb lattice is reduced from $C_{3v}$ to $C_3$ . The width of the bandgap can be modulated dynamically by the electron density of the plasma. The eigenstates of Band 1 and Band 2 at $K$ and $K'$ valleys are shown in figure 5(b) when columns of types 1, 2, and 3 are filled with plasma. The coloured area represents the phase distribution of $E_z$ . The arrow direction represents the Poynting vector direction and the size represents the intensity.

When the Dirac cone is opened by the changes of spatial inversion symmetry, the $K(K')$ points of the band structures will appear to have phases with either left- or right-handed polarization. At this time, there is circularly polarized orbital angular momentum at the valley, which can be described by topological charge. Topological charge is defined by $l=\frac{1}{2\pi}\oint_L \nabla[\arg(E_z)]d\vec{s}$, where $\arg(E_z)$ represents the phase of electric field and $L$ represents the closed path around the singularity of electric field distributions. The right-handed circular polarized (RCP, $l = 1$ ) angular momentum at the $K$ valley and the left-handed circular polarized (LCP, $l = -1$ ) angular momentum at the $K'$ valleys are identified in Band 1(see figure 5(b)). As for Band 2, the LCP angular momentum at the $K$ valley and RCP angular momentum at the $K'$ valleys occurs. When the filled form of the plasma is changed to its opposite, the phase transition occurs as shown in the eigenstate in figure 5(d). The topological charge describes the vortex properties of the eigenmode. The value represents the number of times the phase changes from 0 to $2\pi$ , and the symbol represents the direction of phase change. For example, the phase variation around the electric field singularity for $K$ point of band 1 is shown in figure 5(e) when lower left column is filled with plasma. The horizontal axis represents the position around the electric field singularity, and the vertical axis represents the wrapped phase of $E_z$. It can be seen that the phase positively changes by one cycle along the closed path, which corresponds to a topological charge of +1. At the same time, the direction of the Poynting vector is clockwise, which maintains a better consistency. As for the K point of band 2, the phase negatively changes by one cycle along the closed path corresponding to the topological charge of −1, and the direction of the Poynting vector is counter clockwise as shown in figure 5(b).

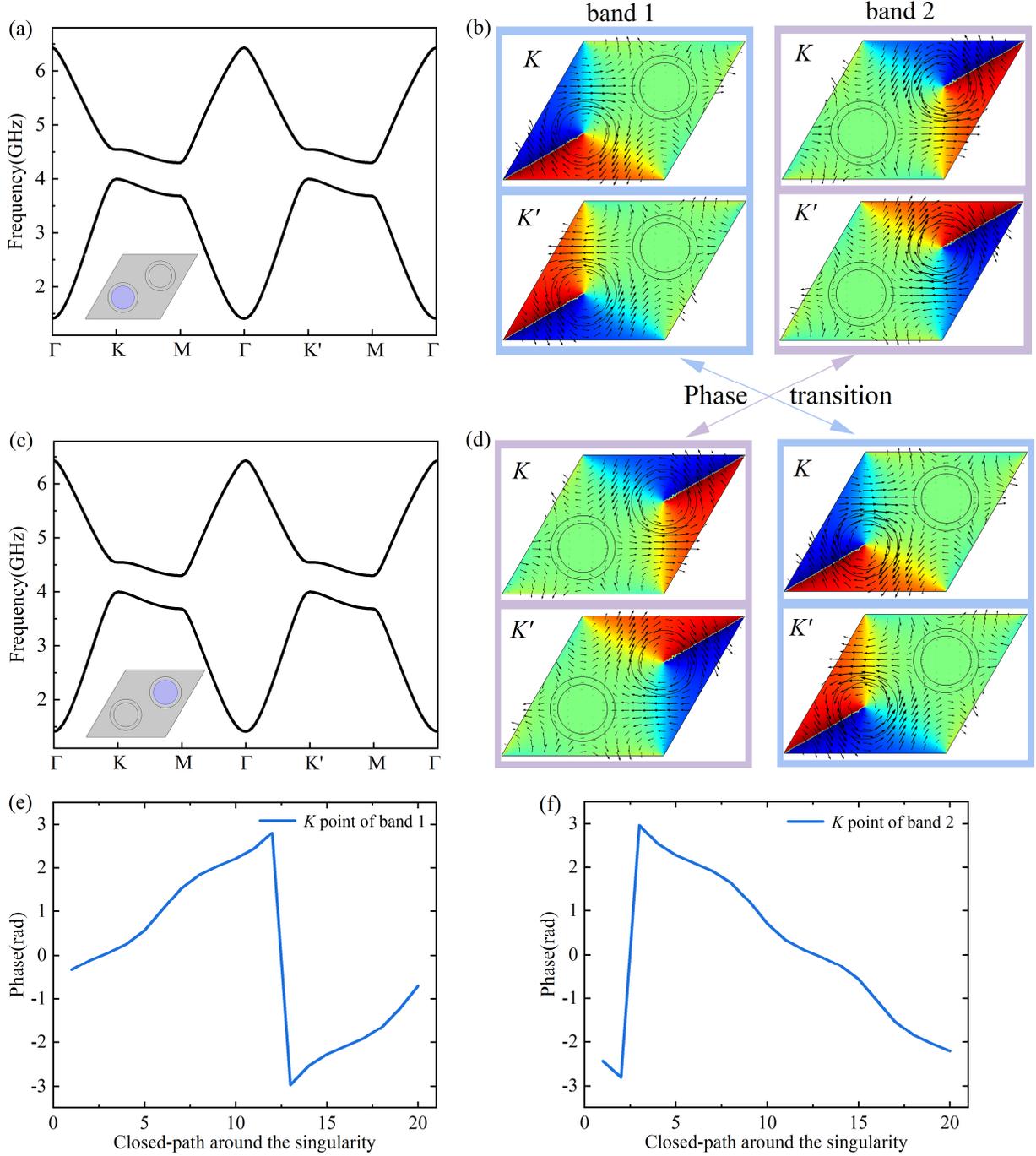

Figure 5. (a) The calculated band structure when plasma fills the lower left column; (b) Phase distribution and Poynting vector at valley K (K') when plasma fills the lower left column; (c) The calculated band structure when plasma fills the upper right column; (d) Phase distribution and Poynting vector at valley K (K') when plasma fills the upper right column; (e) When lower left column is filled with plasma, the phase variation around the singularity for K point of band 1; (f) When lower left column is filled with plasma, the phase variation around the singularity for K point of band 2.

We define structure 1, which has the lower left column in the primitive cell filled with plasma. The structure 2 denotes that the upper right column in the primitive cell is filled with plasma. Splicing structures 1 and 2 together will support a valley-dependent topological edge state. The supercell is constructed as shown in the inset of figure 6 (a), with the upper part being plasma-filled columns of types 1, 2, and 3 and the lower part being plasma-filled columns of types 4, 5, and 6. The Floquet conditions are applied to the boundary in the x-direction of the supercell, and the scattering boundary conditions are used for the boundary in the y-direction. Then, the project band structure is calculated from the supercell (see figure 6(a)). The black line represents the dispersion relationship of the passing bands (bulk states) in the plasma photonic crystal. In addition, the non-bulk state arises in the common bandgap as the blue line. For figure 6(b), we choose 4.16 GHz on the non-bulk state and can see that the electric field is mainly localized at the interface of the two structures, while it decays rapidly in the interior. When the wave vector is fixed to $k_x = 0.22$ (the red point in figure 6(a)), the frequency of the topological edge state is regulated dynamically by the electron density as shown in figure 6(c). The topological properties are still maintained due to the change in electron density that does not cause the phase transition to occur. The eigenfrequency of the non-bulk state increases with the electron density, which expands the scope of application. To excite the valley-dependent topological edge states, point-like chiral sources are constructed by four antennas with the phases set to 0, $\pi/2$, $\pi$, $3\pi/2$. When the phases of the four antennas decreases clockwise with an excitation frequency of 4.16 GHz, electromagnetic energy is transmitted in the left direction along the edge as shown in the $E_z$ distribution of figure 6(d). The path of electromagnetic wave propagation in the backward direction is suppressed. Also, the electromagnetic energy does not leak into the interior of the plasma photonic crystal, which gives the system a high signal-to-noise ratio. However, the electromagnetic wave propagation in the right direction is excited when the phase of the antenna increases clockwise as shown in figure 6(e). The sources with different chirality excite electromagnetic waves

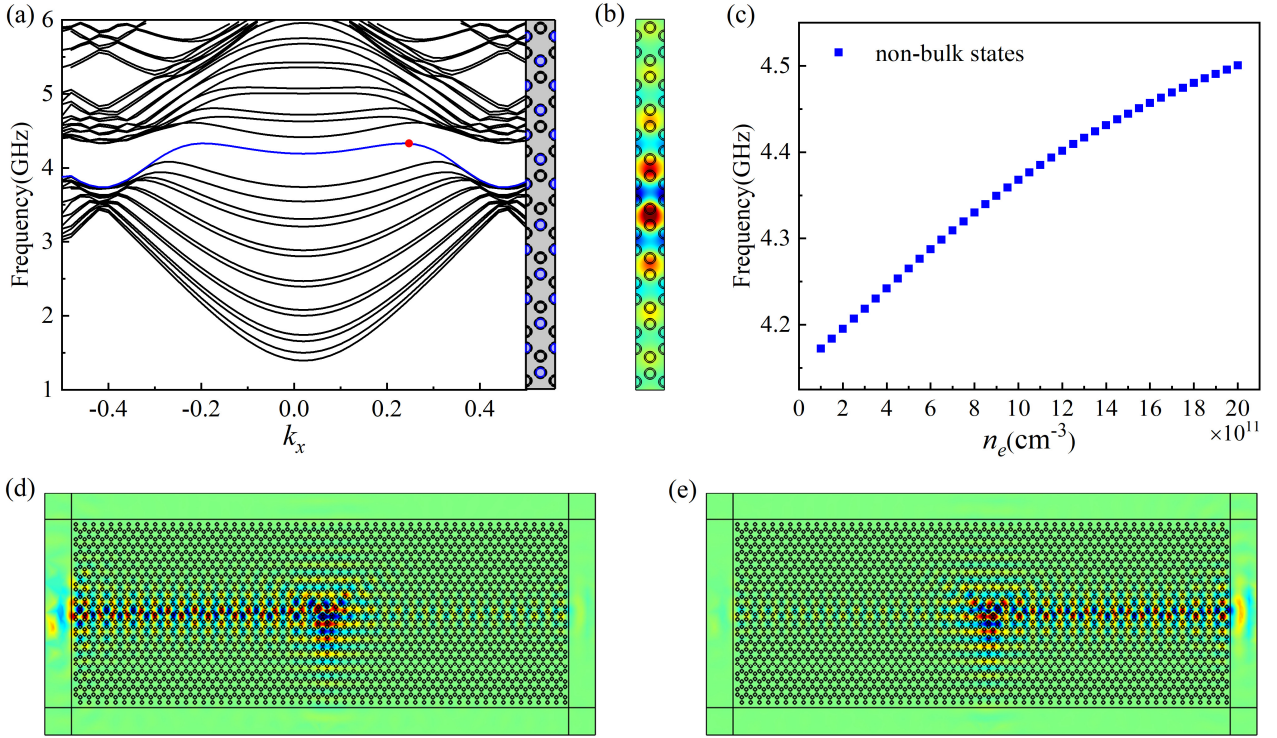

Figure 6. (a) The project band structure of supercell; (b) The z-component of electric field distribution when the frequency of the electromagnetic wave is 4.16 GHz; (c) The non-bulk state varies with plasma electron density; (d) A point-like chiral source with a clockwise decreasing phase is placed at the interface; (e) A point-like chiral source with a clockwise increasing phase is placed at the interface

to propagate in opposite directions. In application areas, when electromagnetic waves propagate through the device, they are strongly reflected when encountering impurities or defects, thus affecting the transmission efficiency. Topological edge states possess the properties of unidirectional propagation, backscatter suppression, and immunity to impurities, which can solve the problem perfectly. Based on the topological edge state theory, topological lasers, optical waveguides, optical isolators, and optical modulators have been developed one after another. The plasma photonic crystal designed in this paper not only realizes topological edge states but also has tunable features, which will have a wide prospective application in tunable microwave devices.

By modulating the electron density in a small range, we obtain tunable topological edge states and the topological properties are preserved. But the results change dramatically when the electron density is extremely high. As is known to us, the photonic crystal with honeycomb lattice has a linear dispersion relation at the $K(K')$ point, which is known as Dirac cone. The Dirac cone has good stability and can maintain its existence even when $\varepsilon_{xx}$ and $\varepsilon_{yy}$ are very different. The topological phase transition must be accompanied by the closing and reopening of the Dirac cone. A commonly method to break the Dirac cone is to apply a magnetic field outside a photonic crystal made of magneto-optical material, which breaks the time-reversal symmetry of the system[38]. Another mechanism is to change the honeycomb lattice to a triangular lattice, which makes the spatial inverse symmetry of the system break. This also causes the Dirac cone to open[16]. In the previous section, the Dirac cone is formed at $K(K')$ point when all tubes are filled with plasma in a primitive cell. When the electron density increases by an order of magnitude, the Dirac cone is opened while the band structure is raised overall (see figure 7(a)). The Dirac point changes when the electron density varies continuously as shown in figure 7(b).

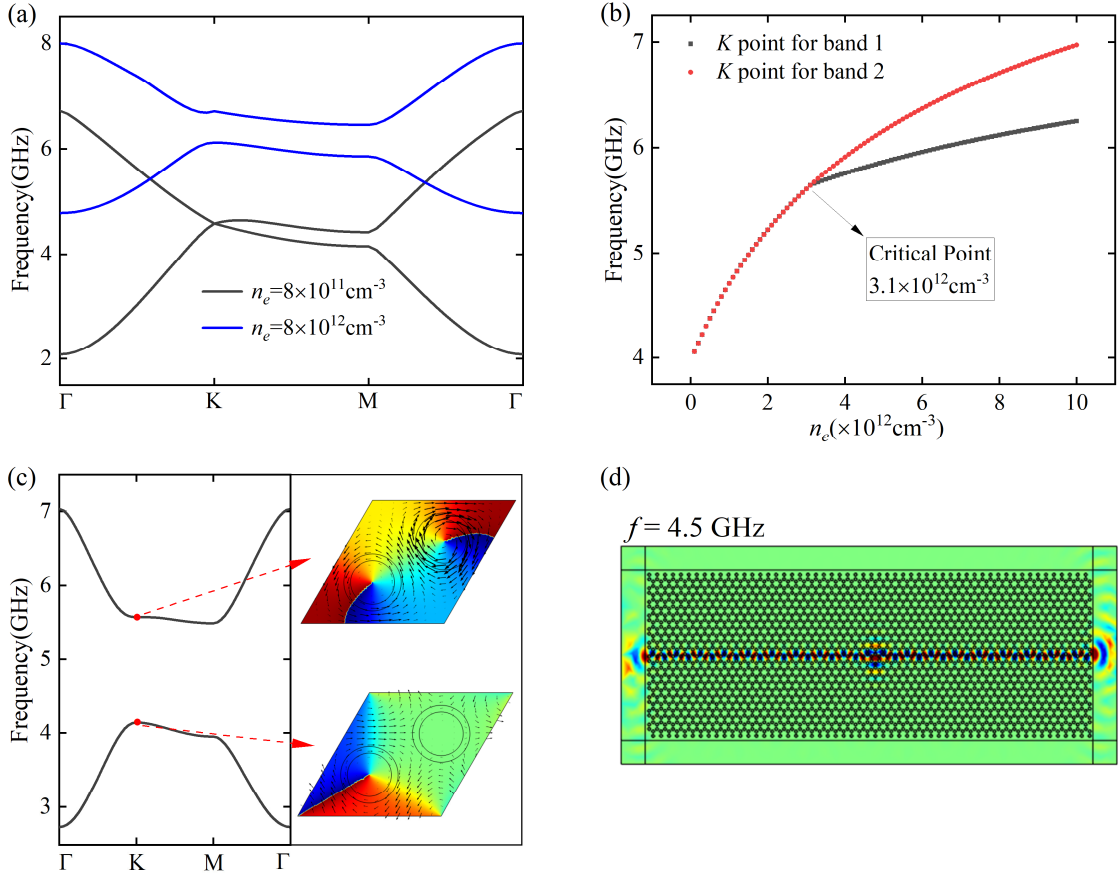

Figure.7 (a) When the electron density is raised to $8 \times 10^{12}$ cm$^{-3}$, the Dirac cone is broken and an omnidirectional band gap appears; (b) The critical density at which the Dirac point is broken; (c) The calculated band structure when plasma fills the lower left column at $8 \times 10^{12}$ cm$^{-3}$; (d) The edge state without topological protection.

From the results, the critical electron density that makes the Dirac cone break is $3.1 \times 10^{12}$ cm$^{-3}$. The permittivity of the plasma at this point is -19.24. Therefore, the negative permittivity can be used as an alternative method to break the Dirac cone independently of applying a magnetic field or changing the lattice structure. However, the band gap obtained by increasing the electron density is not topologically nontrivial. The specific proof is shown in figure 7(c). We increase the electron density of figure 5(a) to $8\times10^{12}$ cm$^{-3}$. It can be seen from the results that the band gap is widened and its topological properties are changed. Since the Poynting vector at point $K$ in the two bands does not have symmetry as in figure 5(b). When the chiral source is placed at the interface, the electromagnetic wave propagates along two directions at an electron density of $8\times10^{12}$ cm$^{-3}$ as shown in figure 7(d). The edge states in this case are defective states and not protected by topology. Although the method of breaking the Dirac cone by increasing the electron density does not lead to a topological phase transition of the system, the effect caused by the negative permittivity deserves to be studied in depth.

## 4. Conclusions

The plasma photonic crystal is the intersection of plasma and photonic crystal disciplines, which both have the properties of photonic crystals and plasma. It can be used for tunable filters and waveguides, etc. by changing the discharge plasma parameters and the applied magnetic field. In past research, the band structures of plasma photonic crystals have been focused on different primitive cells, defect structures, and spatial dimensions, which need to be expanded in terms of theory and application. This paper is devoted to investigating the tunable topological states in plasma photonic crystals that cannot be achieved in all-dielectric photonic crystals.

We experimentally measure the electromagnetic parameters of the discharge plasma by the NRW method and TEM cell. The electron density extracted by the effective medium theory is between $1\times10^{11}$ cm$^{-3}$ and $1\times10^{12}$ cm$^{-3}$, and the collision frequency between electrons and neutral atoms is near $1.5\times10^{10}$ Hz. The precisely measured collision frequencies are much lower than the results in the literature such as $9\times10^{10}$ Hz by excluding the absorption effect of mercury vapor. After that, 28 discharge tubes form a two-dimensional plasma photonic crystal while the transmittance is measured at different discharge currents. The band structure of the plasma photonic crystal was calculated by COMSOL software, demonstrating the band gaps in the ranges of 0 GHz to 1.9 GHz and 4.9 GHz to 5.5 GHz in the direction of wave propagation. The collisionless cold plasma approximation ($\varepsilon_r(\omega)=1-\omega_{pe}^2/\omega^2$) is used in the simulation, and the electron density is $8\times10^{11}$ cm$^{-3}$ as measured by the TEM cell. The experimental results agree very well with the simulation results, which proves the rationality of this approximation. Finally, we design a honeycomb lattice of plasma photonic crystals theoretically. The valley-dependent topological phase transition is achieved by breaking the $C_{3v}$ symmetry whose position of plasma is changed in the primitive cell. The edge states are generated at the interface of two plasma photonic crystals with different valley topological properties, and the direction of wave propagation is controlled by the chiral source. The frequency corresponding to the topological edge states can be dynamically regulated by the plasma parameter. More importantly, there is a critical electron density to break the Dirac cone, which originates from the negative permittivity of the plasma. This is worthy of further study in manipulating the flow of electromagnetic waves.

## Acknowledgments
This work was supported by the Natural Science Foundation of China (NSFC, Grant Nos. 12175050).